\documentclass[twocolumn,secnumarabic,amssymb, nobibnotes, aps, prd,superscriptaddress]{revtex4-2}
\usepackage[british]{babel}

\usepackage{graphicx}
\usepackage{color}
\usepackage{amsmath}
\usepackage{gensymb}
\usepackage{tabularx}
\usepackage{multirow}
\usepackage{notes2bib}
\usepackage[left,modulo]{lineno}
\usepackage{upgreek}

\makeatletter
\newcommand{\maintextlabel}[2]{\def\@currentlabel{#2}\label{#1}}
\makeatother

\begin{document}
\title{Orientational dynamics in supercooled glycerol computed from MD simulations: self and cross contributions}

\author{Marceau H\'enot}
\email[Corresponding author: ]{marceau.henot@cea.fr}
\affiliation{SPEC, CEA, CNRS, Université Paris-Saclay, CEA Saclay Bat 772, 91191 Gif-sur-Yvette Cedex, France.}

\author{Pierre-Michel Déjardin}
\affiliation{Laboratoire de Modélisation Pluridisciplinaire et Simulations, Université de Perpignan Via Domitia, 52 avenue Paul Alduy, F-66860 Perpignan, France.}

\author{François Ladieu}
\affiliation{SPEC, CEA, CNRS, Université Paris-Saclay, CEA Saclay Bat 772, 91191 Gif-sur-Yvette Cedex, France.}

\date{\today}
\begin{abstract}
The orientational dynamics of supercooled glycerol using molecular dynamics simulations for temperatures ranging from 323~K to 253~K, is probed through correlation functions of first and second ranks of Legendre polynomials, pertaining respectively to dielectric spectroscopy (DS) and depolarized dynamic light scattering (DDLS). The self, cross, and total correlation functions are compared with relevant experimental data. The computations reveal the low sensitivity of DDLS to cross-correlations, in agreement with what is found in experimental work, and strengthen the idea of directly comparing DS and DDLS data to evaluate the effect of cross-correlations in polar liquids. The analysis of the net static cross-correlations and their spatial decomposition shows that, although cross-correlations extend over nanometric distances, their net magnitude originates, in the case of glycerol, from the first shell of neighbouring molecules. Accessing the angular dependence of the static correlation allows us to get a microscopic understanding of why the rank-1 correlation function is more sensitive to cross-correlation than its rank-2 counterpart. 
\end{abstract}
\maketitle

\section{Introduction.}
Dielectric spectroscopy (DS) is a powerful tool for studying polar supercooled liquid dynamics~\cite{kremer2002broadband, lunkenheimer_dielectric_2002, Lunkenheimer2018}. The outcome of the measurement, the complex dielectric permittivity $\epsilon(\omega)$, contains a wealthiness of information regarding the collective orientational motion of the permanent dipoles of the constitutive molecules, and more precisely on the relaxation processes at work in the liquid under scrutiny \cite{bottcher_theory_1973,bottcher1978dielectrics}. The broad range of available frequencies ($10^{-5} - 10^{13}$~Hz) for a wide range of temperatures, allows to follow the slow down of the structural $\alpha$ relaxation upon cooling close to the glass transition temperature as well as the emergence of secondary relaxation processes such as JG processes, believed to be an intrinsic characteristic of glassy dynamics~\cite{ngai2004classification}, or the excess wing, recently associated with dynamical facilitation~\cite{guiselin2022microscopic}. DS can also be used to characterize the cooperative nature of the $\alpha$ relaxation~\cite{berthier2005direct}, to determine density scaling of the relaxation time~\cite{alba2002temperature, roland2005supercooled}, or to study physical aging of out-of-equilibrium liquids~\cite{leheny_frequency-domain_1998, hecksher2010physical}. 

The complex dielectric permittivity obtained from DS measurements can be linked to the time-dependent equilibrium ﬁeld free total dipole moment correlation function $C_1(t)$ through a Fourier-Laplace transform~\cite{Klug2003, Rival1969, Scaife1998}
 \begin{eqnarray}
\nonumber
\frac{9k_\mathrm{B}T \epsilon_0}{\rho} \frac{(\epsilon(\omega)-\epsilon_{\infty})(2\epsilon(\omega)+\epsilon_{\infty})}{\epsilon(\omega)(\epsilon_\infty+2)^2}\qquad\qquad\\
 = \mu^2 C_{1}(0) \left(1-i\omega\int_{0}^{\infty}C_{1}(t)e^{-i\omega t}dt\right)
\label{KKVRSeq}
\end{eqnarray}
where $k_\mathrm{B}$ is Boltzmann's constant, $\rho$ is the liquid density, $T$ its temperature and $\epsilon_\infty$ is the permittivity at visible optical frequencies. The dipole correlation function of rank $\ell$ is:
\begin{equation}
    \label{eq_corr_Leg}
    C_\mathrm{\ell}(t) = \frac{1}{N}\left\langle \sum_i\sum_j P_\mathrm{\ell}[\cos \vartheta_{i,j} (t_0, t_0+t) ]\right\rangle_{t_0}
\end{equation}
where $N$ is the number of dipoles in the cavity considered, $P_\mathrm{\ell}$ is the Legendre polynomial of rank $\ell$ and $\vartheta_{i,j}(t_0, t_0+t)$ corresponds to the angle between molecule $i$ at time $t_0$ and $j$ at $t_0 + t$. In DS, this angle is measured between dipole moments and the technique is sensitive to the order $\ell = 1$ leading to:
\begin{equation}
    \label{eq_corr_l1}
    C_\mathrm{1}(t) = \frac{1}{N\mu^2} \left\langle \sum_i\sum_j \vec{\mu}_i(t_0) \cdot \vec{\mu}_j(t_0+t) \right\rangle_{t_0}
\end{equation}
The static value can be rewritten in: 
\begin{equation}
\label{eq_gk}
    C_\mathrm{1}(0) = g_\mathrm{K} = 1 + \frac{1}{N\mu^2}\left\langle \sum_i \sum_{j \neq i}\vec{\mu}_i(t_0) \cdot \vec{\mu}_j(t_0) \right\rangle_{t_0}
\end{equation}
where $g_\mathrm{K}$ is the Kirkwood correlation factor that can either be $>1$, in which case the dipole-dipole correlation are overall positive, or $<1$ meaning that anti-alignments dominates. The dynamics is also expected to be affected by cross-correlation because there is \textit{a priori} no reason that the timescales and shapes of the self and cross-correlations function coincide exactly. A striking example of dynamical consequences of intermolecular correlations is the behavior of mono-alcohols which display another relaxation process at low frequencies, called the Debye peak, related to the formation of supramolecular H-bonded structures consisting of chains ($g_\mathrm{K}>1$) or rings ($g_\mathrm{K}<1$)~\cite{bohmer2014structure}. A recent theory from Déjardin~\textit{et al.}~\cite{dejardin2019linear}, showed that the liquid dynamics can be strongly affected by the effect of positive cross-correlations.

Recently, results from DS were compared to other techniques less sensitive to cross-correlations. The fluorescence response of a local probe diluted in a mono-alcohol was shown to be insensitive to the Debye relaxation of the liquid, allowing it to disentangle it from the other relaxation processes~\cite{weigl2019local}. Another technique that has been proven useful in that regard is depolarized dynamic light scattering (DDLS) which probes molecular orientations through the anisotropy of the polarizability. The relevant correlation function is given by eq.~\ref{eq_corr_Leg} with $\ell=2$. It follows that the technique does not distinguish between parallel and antiparallel alignments. There is strong experimental evidence that DDLS is insensitive to cross-correlations. For example, Gabriel~\textit{et al.}~\cite{gabriel2017debye} showed that in mono-alcohols DDLS displays an $\alpha$ peak but no Debye peak. In addition, in a non-associating liquid, Pabst~\textit{et al.}~\cite{pabst2020dipole} showed that progressively diluting the system in a non-polar solvent leads the DS spectra to look more and more alike the DDLS spectra. All of this illustrates the importance of cross-correlation effects in DS which can significantly broaden the $\alpha$ peak. Moreover, while the shape of the $\alpha$ peak in DS spectra is system dependent, it was shown in DDLS to follow a generic line shape of slope $-1/2$ on the high-frequency flank~\cite{pabst2021generic}. There is still debate, however, on whether this generic response reflects the true structural relaxation better than the dielectric one~\cite{moch2022molecular}.

When dealing with physical processes taking place at the nanometric scale, molecular dynamics (MD) simulation is an attractive method that can give access to microscopic observables that are otherwise hard, or impossible, to obtain experimentally. This method is however limited to high temperatures or simplified systems, due to its computational cost. To study the generic behavior of liquid glass-formers, model systems can be thought of being made of polydisperse beads interacting through a Lennard-Jones potential. This helped give information on the spatio-temporal nature of relaxations~\cite{stein_scaling_2008, guiselin2022microscopic}. Another approach, more suitable for direct comparison with experiments, is to rely on a more precise modelization of specific molecules, taking into account their dipolar nature and electrostatic interactions. This gives access to their dielectric response~\cite{edwards_computer_1984, saiz_dielectric_2000, zhang2016computing, atawa2019molecular, olivieri2021confined}. Recently, MD simulations on a model dipolar system showed that, while the orientational $\ell = 1$ correlation function of weakly polar systems is dominated by the self response, strongly polar liquids are much affected by cross-correlations~\cite{koperwas2022computational}.

The wide variety of organic liquids available has led to the choice of some systems, considered as models or representatives. Glycerol, by its apparent simplicity and its low tendency to crystallize, has long been the subject of extensive studies, by various techniques including dielectric spectroscopy~\cite{davidson1951dielectric, leheny_frequency-domain_1998, schneider1998dielectric, kudlik1999dielectric, lunkenheimer_dielectric_2002}, neutron spectroscopy~\cite{wuttke1996structural}, nuclear magnetic resonance (NMR)~\cite{meier2012intermolecular}, DDLS~\cite{brodin2005depolarized, gabriel_intermolecular_2020} and MD simulations~\cite{chelli1999glycerol, chelli_glycerol_1999_2, blieck_molecular_2005, egorov_molecular_2011, busselez2011non, busselez2014structural, seyedi2016dynamical, becher2021molecular}. Its dynamics is, however, not particularly simple. As a tri-alcool, it is subject to H-bonds but does not display a Debye peak that would result from linear supramolecular chains. Shear mechanical spectroscopy has shown the existence of a low-frequency mode that is believed to result from the hydrogen-bonded network formed between molecules~\cite{jensen2018slow}.

In this article, we report a MD study of the orientational dynamics of glycerol, on a large temperature range (from 253 to 323~K) reaching the moderately supercooled regime, simulated from a model already widely used in the literature~\cite{chelli1999glycerol, chelli_glycerol_1999_2, blieck_molecular_2005, egorov_molecular_2011, becher2021molecular} over durations of up to 7 $\upmu$s. We first compute the self response of the dipolar moment for ranks $\ell = 1$ and 2 from which we deduce the loss function $\chi^{\prime\prime}_{\ell}(f)$ for frequencies down to 200~kHz. We then analyze the cross-correlation and we exploit the possibility offered by MD to decompose this part of the response as a function of the relative distance and orientation of the dipoles. We compute the total loss function for both ranks as well as the part resulting from cross-correlations alone which allowed us to verify that cross-correlations play a major role in the $\ell=1$ response while being almost negligible for $\ell=2$. We compare these data to experimental DS and DDLS spectra and obtain similar temperature dependence for the relaxation time and slope of the high-frequency flank of the $\alpha$ peak. We discuss how the differences in the spectra associated with different ranks can be related to the underlying molecular relaxation mechanisms. Moreover, we show that, for glycerol, the net cross-correlation originates only from the first shell of neighouring molecules. This is the case for both $\ell=1$ and 2 although their different sensitivity to orientational correlations leads, at the end, to significant differences in the importance of contributions coming from cross-correlations.

\section{Methods}
The molecular dynamics (MD) simulations were performed using OpenMM~\cite{openmm_2017} on an Nvidia RTX A5000 GPU. Glycerol has been modeled using the re-parameterized AMBER force field previously employed in the literature~\cite{chelli1999glycerol,chelli_glycerol_1999_2,blieck_molecular_2005,egorov_molecular_2011,busselez2011non, busselez2014structural, becher2021molecular} and whose parameters are given in the suppl.
mat. Atoms belonging to the same molecule interact through harmonic potentials for bond length and angle and a periodic potential for bond torsion. Non-bounded atoms interact through a Lenard Jones potential with a 1~nm cutoff and a coulomb interaction computed using a Particle Mesh Ewald (PME) algorithm (1~nm cutoff and 0.0005 error tolerance). The simulation does not account for electronic polarizability. Each atom carries a constant partial charge originally derived by Chelli~\textit{et al.}~\cite{chelli1999glycerol} from quantum mechanical calculations. Later, Blieck~\textit{et al.}~\cite{blieck_molecular_2005} noticed that this parameterization led, in the temperature range 333 - 413 K, to a dynamics 10 times faster than measured experimentally by neutron spectroscopy. They slowed down the dynamics by the right amount by reducing by 5~\% the hydroxyl group atomic charges. They also checked that the simulation reproduced fairly well the static structure factor measured by neutron scattering~\cite{champeney1986structural}.
This corresponds to a mean dipole moment of $\langle \mu \rangle = 3.2$~D  which is higher than the $\mu_\mathrm{exp} = 2.68$~D value measured in a nonpolar solvent~\cite{rizk1968dipole}. This can be seen as a way to compensate for the absence of electronic polarizability which leads, in the real system, through the reaction field, to an effective dipole moment greater than $\mu_\mathrm{exp}$~\cite{bottcher_theory_1973}.
The same parameters were later used by Egorov~\textit{et al.}~\cite{egorov_molecular_2011} (who corrected slightly the charges to ensure the molecule neutrality, and made all bound flexible) to study glycerol-water mixtures and more recently by Becher~\textit{et al.}~\cite{becher2021molecular} to reproduce NMR spectra in the 300-540~K range. The parameters used in this work were almost identical with only small modifications intended to reduce the computational cost: the length of bonds involving hydrogen were fixed (as in refs.~\cite{chelli1999glycerol, blieck_molecular_2005}) and hydrogen atom mass was increased by 40\% allowing to use an integration time of 4~fs. The simulations were carried out on a system of $N=2160$ molecules (30 240 atoms) in a cubic cell of side length $a \approx 65$~\AA~ with periodic boundary conditions (PBCs), in the NPT ensemble at eight different temperatures $T$ (from 323 to 253~K) and at pressure $P=1$~bar using a Monte Carlo barostat and a Nosé-Hoover thermostat. In order to study the effect of the system size, a simulation at $T=323$~K was performed on a system consisting only of $N=540$ molecules ($a \approx 41$~\AA). Random initial states were generated using Packmol~\cite{packmol_2009}, equilibrated at $323$~K and progressively cooled down to 253~K by 10~K steps by waiting at each step an equilibration time corresponding to 98 to 200~$\tau_\alpha$, reaching 7~$\upmu$s (see details in the suppl. mat.). At each temperature, simulation runs lasted from more than 180~$\tau_\alpha$ for $T\geq 263$~K and 67~$\tau_\alpha$ at 253~K (corresponding to 4~$\upmu$s). For all simulation runs, the dipole of each molecule $\vec{\mu}_{i}(t)$ ($i\in [1,N]$) and its position $\vec{r}_{i}(t)$ were determined by computing the barycenter of the positive ($q_{+}$ at $\vec{r}_{+}$) and negative charges ($q_{-} = - q_{+}$ at $\vec{r}_{-}$) with $\vec{\mu} = q_{+}(\vec{r}_{+}-\vec{r}_{-})$ and $\vec{r} = (\vec{r}_{+}+\vec{r}_{-})/2$. 

\section{Results and discussion}

\subsection{Self correlation} 
\begin{figure}[htbp]
  \centering
  \includegraphics[width=8.9cm]{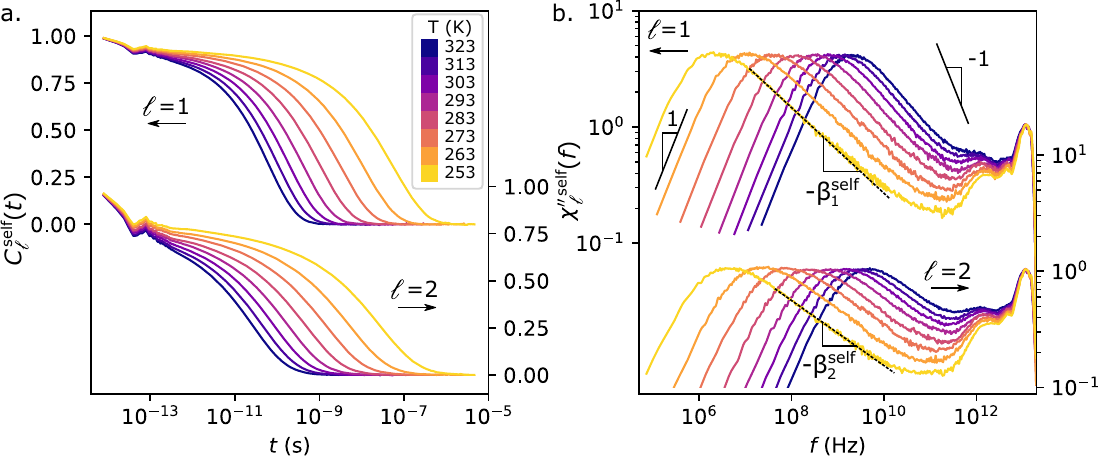}
 \caption{(a) Dipole self correlation functions for the Legendre polynomial of rank $\ell = 1$ (top) and $\ell = 2$ (bottom) at different temperature $T$ ranging from 323 K to 253 K with 10~K steps. (b) Dielectric loss function corresponding to the self part of the correlation functions $\ell = 1$ and 2. The black dashed lines correspond to a power law fit of slope $-\beta^\mathrm{self}_{\ell}$ on the high-frequency wing.}
  \label{fig1}
\end{figure}

The self dipole correlation function is defined by:
\begin{equation}
\label{eq_C_self}
    C^\mathrm{self}_{\ell}(t) = \left\langle P_\ell [\cos \vartheta_{i,i} (t_0, t_0+t)]  \right\rangle_{i, t_0}
\end{equation}
It characterizes the molecular relaxation through rotational movement of the permanent dipole. This function is shown in fig.~\ref{fig1}a at each temperature, for rank $\ell=1$ and 2. Three regimes can be observed: at short time ($t<100$~fs) a small decorrelation occurs and a boson peak is visible at $t \approx 70$~fs. At long time, there is a complete, non-exponential decorrelation ($C^\mathrm{self}_\ell(t)$ reaches 0) corresponding to the $\alpha$ relaxation. At intermediate times the correlation is high but slowly decreasing. This regime is almost nonexistent at 323~K but extends over two decades at $T=253$~K. While the global shape is the same for $\ell = 1$ and 2, the short time decorrelation appears more intense for $\ell = 2$. This is simply due to the quicker decreases of $P_2(\cos \vartheta)$ compared to $P_1(\cos \vartheta)$ for $\vartheta \ll 1$.

The mean self relaxation time is obtained from $\tau^\mathrm{self}_\ell = \int_0^\infty C^\mathrm{self}_\ell(t)\mathrm{d}t$ and is shown as a function of $1/T$ in blue in fig.~\ref{fig6}a. The relaxation times are shorter for $\ell = 2$ (empty markers) than for $\ell = 1$ (solid markers) and they both display a super-activated behaviour.

The self loss function was obtained, following eq.~\ref{KKVRSeq}, by applying the fluctuation-dissipation theorem~\cite{livi2017nonequilibrium}: $\chi^{\prime \prime \mathrm{self}}_\ell(f) \propto f \times \mathrm{TF}(C^\mathrm{self}_\ell(t))$ where $\mathrm{TF}$ is the Fourier transform, computed using the fftlog algorithm adapted to log spaced data~\cite{Hamilton_2000}. The fact that the correlation function was averaged on long times ($\approx 100 \tau_\alpha$) leads to a fairly low amount of noise on the spectra, shown in fig.~\ref{fig1}b. They were all rescaled by superimposing their microscopic peaks at $10^{13}$~Hz. The frequency at which the maximum of the $\alpha$ peak is reached was found to correspond (within the uncertainty) to $1/(2\pi  \tau^\mathrm{self}_\ell )$. On the low-frequency side, the spectra follow a power law with slope 1, as expected. On the high-frequency flank of each spectrum, there is a power law regime on one to two decades in frequency with a slope $-\beta^\mathrm{self}_\ell$, interrupted by the fast process~\cite{Lunkenheimer2018}. The corresponding values of $\beta^\mathrm{self}_\ell$ are shown in blue in fig.~\ref{fig6}b. For $\ell = 1$ (solid markers), slope increases with temperature (ranging from 0.36 at 253~K to 0.46 at 323~K) while for $\ell = 2$ (empty markers), it is temperature independent and systematically smaller ($\approx 0.27$). These low $\beta$ values are associated with the non-exponential nature of the relaxation process.

\subsection{Static cross-correlation} 
\label{section_static_cross}

As stated in the introduction, experimental methods such as DS and DDLS are sensitive, not only to the self correlation function, but rather to a total correlation made of the self part and of a cross-correlation part. We thus need to get access to the correlation function associated to cross-correlation:
\begin{equation}
\label{eq_C_cross}
    C^\mathrm{cross}_{\ell}(t) = \frac{1}{N}\left\langle \sum_i \sum_{j \neq i} P_\ell [\cos \vartheta_{i,j} (t_0, t_0+t)] \right\rangle_{t_0}    
\end{equation}

However, one has to be careful with the application of this definition directly to the MD simulation box due to the effect of PBCs on the treatment of electrostatic interactions. With the PME method used here, our simulation box can be seen as wrapped in tinfoil, or embedded in an infinite medium in which the macroscopic electric field is null~\cite{caillol1992asymptotic, zhang2016computing, olivieri2021confined}. This effect is responsible for a long-range dipole correlation of significant amplitude, that is an artifact of the simulation, and which cannot be suppressed or diminished by increasing the simulation box size (see fig.~\ref{suppl_fig_box} of suppl. mat.). This artificial cross-correlation is maximum on average for couples of molecules separated by a distance of the order of the box size $a$. A way to get around this difficulty is to use a simulation box large enough to decouple the real correlation (occurring at relatively small distances) and the artifact~\cite{zhang2016computing, olivieri2021confined}.

\begin{figure}[htbp]
  \centering
  \includegraphics[width=8.9cm]{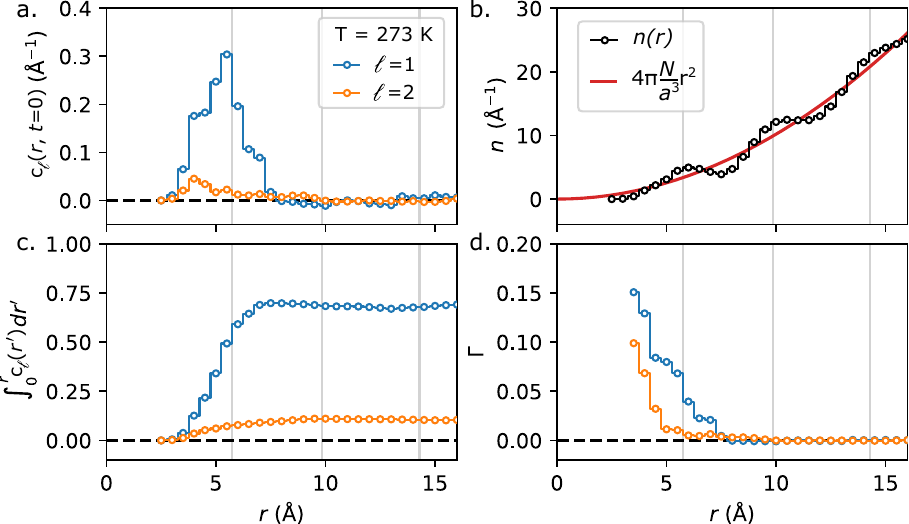}
 \caption{Distance dependence of the dipole static cross-correlation function ($\ell = 1$ and 2) at $T = 273$~K. (a) Contribution $c$ per unit distance to the cross-correlation of all the molecules situated at $r$ (b) Number of molecules per unit distance at $r$. The red curve corresponds to a homogeneous medium of the same density. The vertical grey lines on all the plots correspond to the first, second and third neighbour peaks. (c) Contribution to the cross-correlation of all the molecules within a sphere of radius $r$. (d) Mean level of cross-correlation $\Gamma$ ($\in [-1,1]$) for the molecules at $r$.}
  \label{fig2}
\end{figure}
We decompose the cross-correlation function $C^\mathrm{cross}_\ell(t)$ into contributions per unit distance denoted $c_\ell(r, t)$ depending on the distance $r = \|\vec{r}_{j} - \vec{r}_{i} \|$ between the reference molecule $i$ and all molecules within $[r, r+\mathrm{d}r]$. This quantity, computed for slices of 0.5~\AA~ and averaged over $i$ and $t_0$, is plotted in fig.~\ref{fig2}a for the static case ($t=0$) for $\ell = 1$ and 2. The dipole density $n(r)$ is plotted in black in fig.~\ref{fig2}b alongside with the parabola in red that would be obtained for a homogeneous system of the same average density. The difference between these two curves shows a series of maxima that corresponds to the first, second and third neighbour peaks. For $\ell = 1$, it appears that the cross-correlation contribution $c_1(r)$ (in blue) is maximum for the first neighbours and reaches zero before the second layers of neighbours. Fig.~\ref{fig2}c represents,  as a function of $r$, the total contribution $\int_0^r c(r^\prime)\mathrm{d}r^\prime$ integrated within a sphere of radius $r$. We see that the cross-correlation reaches a plateau at $r \approx 7$~\AA~ while the cross-correlation coming from PCBs starts to be perceptible for $r>20$~\AA~ (see suppl. mat.). It is also interesting to study the mean level of static cross-correlation $\Gamma(r) = c(r)/n(r)$, plotted in fig.~\ref{fig2}d that shows that the cross-correlation per dipole is positive, is a strictly decreasing function of the distance and is only of the order of 5-15~\% in average for the nearest neighbours. From these data, we can deduce the Kirkwood factor $g_\mathrm{K} = 1 + \int_0^{r_\mathrm{lim}} c(r^\prime)\mathrm{d}r^\prime = 1.70 \pm 0.02$ at 273~K, with $r_\mathrm{lim} = 7.5$~\AA. This is smaller than the $2.6 \pm 0.2$ value reported in the literature and deduced from static permittivity measurements~\cite{gabriel_intermolecular_2020} but it is compatible with previous numerical results on glycerol for $T>250$~K~\cite{seyedi2016dynamical}. It is also interesting to note that our value of $\mu^2 g_\mathrm{K}$ (which is the quantity accessible from the experiments, see eq.~\ref{KKVRSeq}) matches exactly the experimental value and displays the same temperature dependence (cf fig.~\ref{suppl_mu2_gk} in suppl. mat.). In other words, the simulation gives the expected value of $\epsilon(0)$ in the whole temperature range. However, as the dipolar moment of glycerol has been measured with a reasonable accuracy in a non-polar solvent~\cite{rizk1968dipole}, it is likely that the MD simulation underestimate the value of $g_\mathrm{K}$. For $\ell = 2$, the cross-correlation (in orange) appears to be also due to the first layer of neighbours but is much less intense than for $\ell = 1$ and its integrated value saturates for a slightly higher radius. With $r_\mathrm{lim} = 9.5$~\AA, the quantity analogous to the Kirkwood factor for $\ell = 2$ would be only $1.11 \pm 0.01$.

\begin{figure}[htbp]
  \centering
  \includegraphics[width=8.6cm]{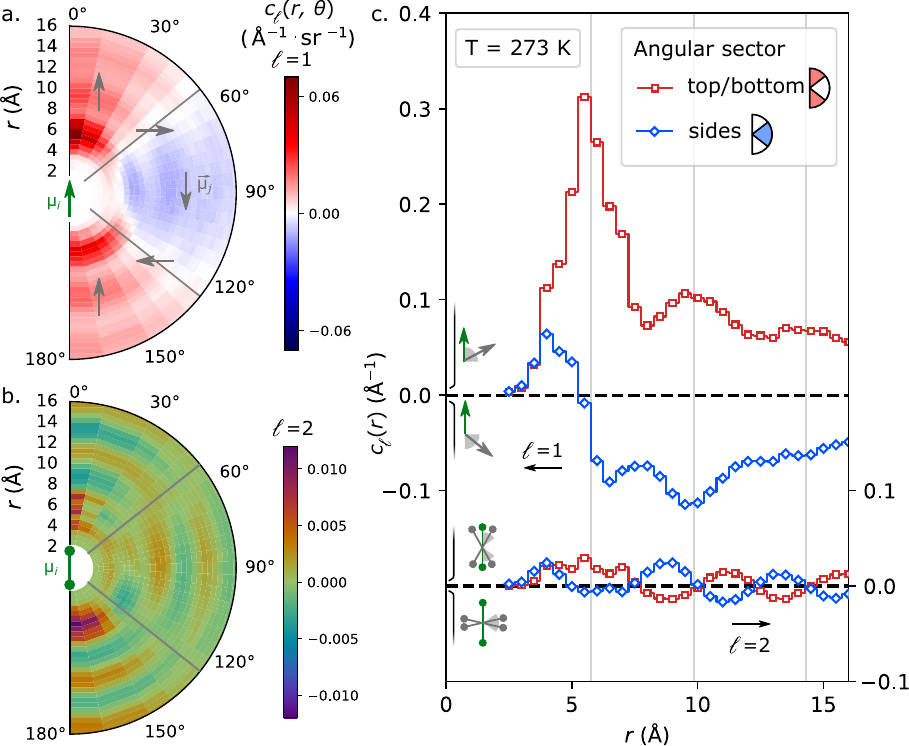}
 \caption{Relative orientation and distance dependence of the static dipole cross-correlation functions at $T = 273$~K. (a) Contribution $c$ per unit distance and solid angle to the cross-correlation ($\ell = 1$) of all the molecules situated at $r$ and at angle $\theta$. Red zones are positively correlated while blue zones are anti-correlated as illustrated by the grey dipole drawings. (b) Same plot than a. for $\ell=2$. (c) Same quantity as in fig.~\ref{fig2}a, for $\ell=1$ (top) and $\ell=2$ (bottom), but with distinguished contribution from the red sector ($|\cos\theta| > 0.62$, see grey lines on a,b) and from the complement blue sector. Drawings close to the vertical axes illustrate the physical meaning of the correlation sign for $\ell = 1$ (right) and $\ell=2$ (left).}
  \label{fig3}
\end{figure}
The dipolar interaction is anisotropic in nature and it makes sense, rather than averaging the cross-correlation over all dipoles situated at a given distance, to distinguish the contribution as a function of the relative orientation. Spherical coordinates $(r, \theta, \phi)$ can be defined with respect to the reference dipole $i$ where $\theta$ is the angle between $\vec{\mu}_i$ and $\vec{r} = \vec{r}_j-\vec{r}_i$. By symmetry, the contribution should not depend on the azimuthal angle $\phi$. The cross-correlation contribution $c_\ell(r,\theta)$ per unit distance and solid angle is shown in the static case in fig.~\ref{fig3}a ($\ell = 1$) and b ($\ell =2$) as a function of $r$ and $\theta$. For $\ell = 1$, similarly to a recent observation in water~\cite{olivieri2021confined}, the spatial distribution of cross-correlation appears strikingly different than the $\theta$ averaged curves shown previously. The correlation is positive in an angular sector situated above and below the dipole of reference $|\cos \theta| < \cos \theta_\mathrm{lim}$ and mostly negative on its sides. It is null on lines of constant $|\cos \theta| = \cos\theta_\mathrm{lim}$ shown in grey with $\theta_\mathrm{lim} = 52$°. The cross-correlation contributions summed over these two angular sectors are shown in fig.~\ref{fig3}c (top). These positive and negative contributions extend way over what is visible when considering their sum and, while decreasing, are far from negligible at $r = 16$~\AA. However, these contributions cancel each other for $r>7.5$~\AA. This angular dependence of the cross-correlation is what is expected when considering the energy interaction between two electrostatic dipoles as a function of their relative orientation~\cite{maitland1981intermolecular} and drawings of the most favorable dipoles orientations are shown in fig.~\ref{fig3}a. The net contribution of cross-correlation comes from the first shell of neighbouring molecules. Close molecules situated above and below are strongly positively correlated (even more above than below) with an alignment rate reaching 50~\%. This correlation is favored by the dipole-dipole interaction although the top-bottom asymmetry illustrates that at such a close distance, it is not the only interaction playing a role. On the sides, molecules belonging to the first shell but further than $5.2$~\AA~(corresponding to the first neighbor peak) are on average anti-aligned (again as favored by the dipole-dipole interaction), although not enough to compensate for the positive correlation. Finally, side molecules closer than $5.2$~\AA~are positively aligned. This means that there exists other effects (that may be related to constraints constraints on molecule conformation or to the presence of H-bonds) able to compensate for the \textit{a priori} unfavorable situation of having barycenters of the same sign charges facing each other on average.

For $\ell = 2$ also, a complex spatial dependence of the cross-correlation is visible in fig.~\ref{fig3}b. As this quantity is not sensitive to the correlation sign, it shows a very different behavior than $\ell = 1$. Some oscillations, that correlate well with the density inhomogeneity, are visible. Those also appear for $\ell = 1$ but dominate here. After the first layer of neighbours, the contribution from the two angular sectors are in anti-phase and cancel each other on average. Similarly to the $\ell = 1$ case, the net contribution comes from the first neighbour shell but it is interesting to observe, that its different sensitivity to orientational correlations makes this quantity significantly less sensitive to cross-correlations.

\subsection{Global dipole dynamics}
\begin{figure}[htbp]
  \centering
  \includegraphics[width=8.6cm]{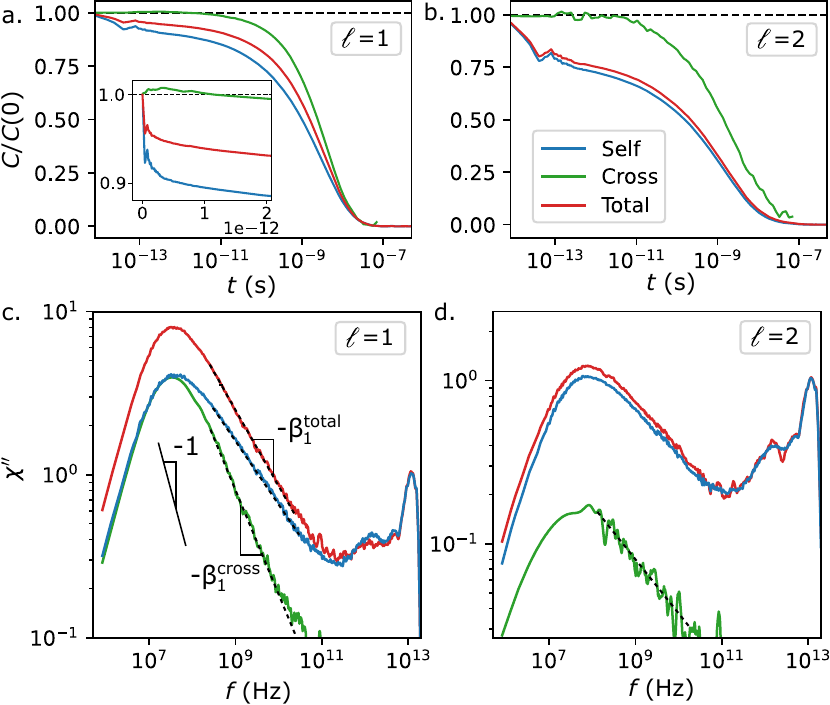}
  \caption{Correlation and loss functions at $T=273$~K. (a, b) Normalized self, cross and total correlation function for $\ell = 1$ (a) and 2 (b). The inset of (a) is a zoom at short time with a linear scale. (c, d) Corresponding loss functions for $\ell = 1$ (c) and 2 (d). The dashed lines in (a) defines the slopes $\beta^\mathrm{cross}_\ell$ and $\beta^\mathrm{tot}_\ell$ on the high frequency flank.}
  \label{fig4}
\end{figure}

For the reason mentioned above and related to the effect of PBCs, the cross-correlation function $C^\mathrm{cross}_\ell(t)$ is computed in the following from eq.~\ref{eq_C_cross} by considering only the molecules located within a sphere of radius $r_\mathrm{lim}$ rather than in all the simulation box. The resulting normalized cross-correlation function is shown in green in fig.~\ref{fig4}a ($\ell=1$) and b ($\ell=2$) at $T=273$~K. The self part is shown in blue and the total correlation function $C^\mathrm{tot}_\ell(t) = C^\mathrm{self}_\ell(t) + C^\mathrm{cross}_\ell(t)$ is shown in red. The cross part does not display a short time decorrelation but for $\ell = 1$, a small amplitude peak is visible at short time ($t<1$~ps) (see inset of fig.~\ref{fig4}a). The mean cross $\tau^\mathrm{cross}_\ell$ and total $\tau^\mathrm{tot}_\ell$ relaxation time, computed by integrating the corresponding normalized correlation functions, are shown for all temperatures in green and red in fig.~\ref{fig6}a. They follow $\tau^\mathrm{self}_\ell < \tau^\mathrm{tot}_\ell < \tau^\mathrm{cross}_\ell$ but with only a 15~\% increase on average between the self and total mean relaxation times for $\ell=1$ and a 25~\% increase for $\ell=2$. This means that the Kivelson and Madden relationship~\cite{kivelson1975theory} ($\tau^\mathrm{tot} = g_\mathrm{K} \tau^\mathrm{self}$) does not seem to be verified in glycerol contrary to other systems such as water~\cite{samanta2022nonlinear}.

\begin{figure}[htbp]
  \centering
  \includegraphics[width=8.9cm]{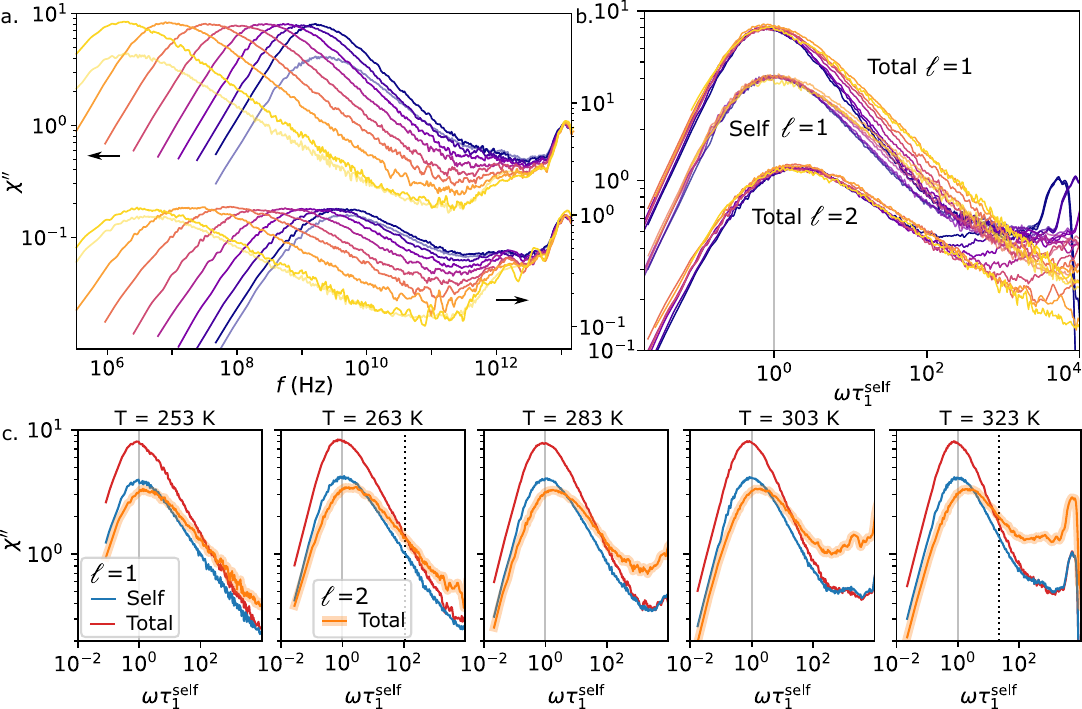}
\caption{Orientational loss function for all temperatures (same color scale than in fig.~\ref{fig1}) studied as a function of $f$ (a) and $\omega \tau_1^\mathrm{self}$ (b). The total loss function is shown for $\ell = 1$ (top, left scale in a) and $\ell = 2$ (bottom, right scale in b). The self part for $\ell = 1$ is shown in (a) only for the extreme temperatures and for all temperatures in (b) (middle). (c) For several temperatures, $\ell = 2$ total spectra (orange) plotted with a scale factor (chosen to correspond to the one determined experimentally by Gabriel~\textit{et al.}~\cite{gabriel_intermolecular_2020}) alongside with the $\ell = 1$ self (blue) and total (red) spectra. The vertical dashed lines shows where the crossing between DS and DDLS data occurs on experimental data.}
  \label{fig5}
\end{figure}

 In the same way as for the self part, a loss function $\chi^{\prime \prime}(f)$ can be computed from the cross part and for the total correlation. The results are plotted for 273~K in fig.~\ref{fig4}c ($\ell=1$) and d ($\ell=2$). It is well visible that the amplitude of the cross-correlation is of the same order as the self part for $\ell =1$ while it is much smaller for $\ell = 2$. On each spectrum, the high frequency flank of the $\alpha$ peak has a power law slope $-\beta$ such that $\beta^\mathrm{self}_\ell < \beta^\mathrm{cross}_\ell < 1$. The slope of the total loss function $\beta^\mathrm{tot}_\ell$ results from both previous slopes as well as the strength of the cross-correlation and takes an intermediate value. The total spectra for all temperatures are shown in fig.~\ref{fig5}a and the values of the slopes $\beta$ are shown in fig.~\ref{fig6}b as a function of $T$: $\beta^\mathrm{tot}_{1}$ is increasing with $T$ while $\beta^\mathrm{tot}_{2}$ appears constant close to $0.3$. This is already visible on the spectra of fig.~\ref{fig5}a but it is even clearer in fig.~\ref{fig5}b where the spectra are shown as a function of a dimensionless frequency $\omega \tau^\mathrm{self}_{1}$. The collapse of the high-frequency side is much better for the total spectra with $\ell=2$ (although the $\alpha$ peak slightly broadens upon cooling) than for $\ell = 1$ (self or total).

\subsection{Comparison with DS and DDLS experiments}
\label{section_comparison_exp}
\begin{figure}[htbp]
  \centering
  \includegraphics[width=8.9cm]{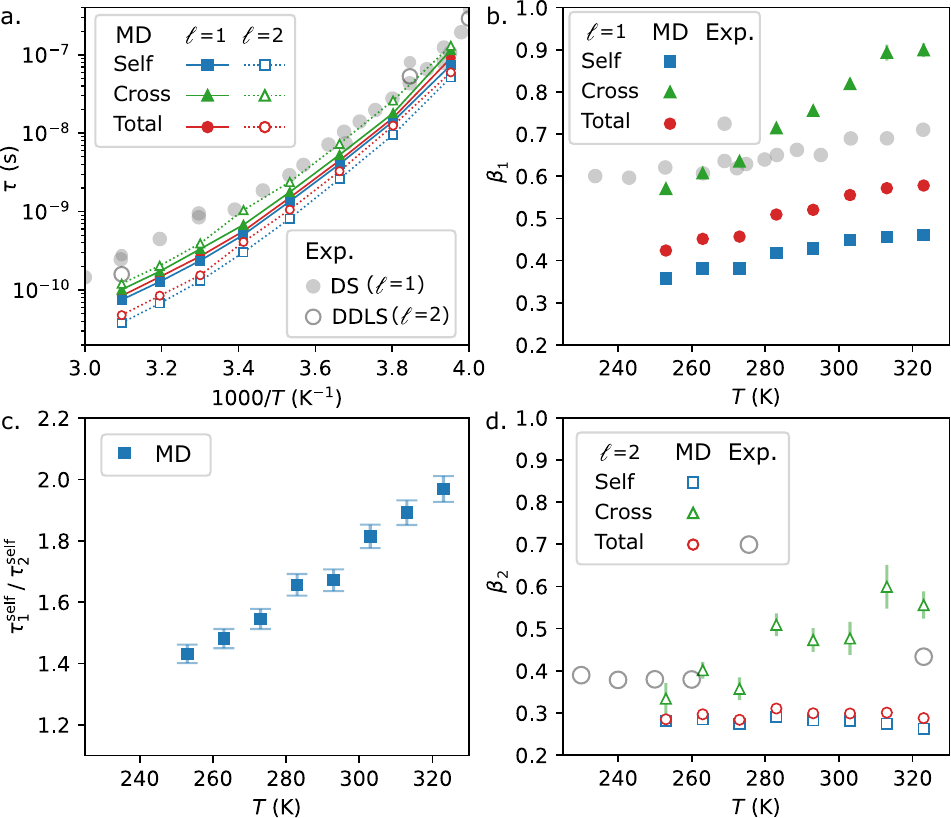}
 \caption{(a) Mean relaxation time obtained by integrating the self (blue squares), cross (green triangles) and total (red circles) dipole correlation function for $\ell = 1$ (solid markers) and $\ell =2$ (empty markers). Markers are linked to improve readability. Relaxation time measured experimentally by DDLS (empty circles) from refs.~\cite{brodin2005depolarized, gabriel_intermolecular_2020} and DS (solids circles) from refs.~\cite{lunkenheimer_dielectric_2002, schneider1998dielectric, gabriel_intermolecular_2020} are shown in grey. (b, d) Power law exponent $\beta$ of the high-frequency wing of the loss function for $\ell = 1$ (b) and $\ell=2$ (d). The color code is the same as for (a). (c) Temperature dependence of the ratio of self mean relaxation time for $\ell= 1$ and $\ell = 2$. In b. and d. experimental data, from refs.~\cite{schneider1998dielectric, gabriel_intermolecular_2020} for DS and refs.~\cite{brodin2005depolarized, gabriel_intermolecular_2020} for DDLS are shown in grey.}
  \label{fig6}
\end{figure}

The spectra $\chi^{\prime\prime~\mathrm{tot}}_{1}(f)$ of fig.~\ref{fig5}a can be compared to experimental glycerol dielectric spectra. The general allure corresponds well to the measurements of Lunkenheimer \& Loidl~\cite{lunkenheimer_dielectric_2002} with a clear $\alpha$ peak showing no excess wing in this range of temperature and a small Boson peak around $10^{12}$~Hz. The temperature dependence of the $\alpha$ relaxation time is compared in fig.~\ref{fig6}a, with DS data as grey solid circles and corresponding MD data $\tau^\mathrm{tot}_{1}$ in red. They show the same trend and could be fitted with a WFL law with close parameters (up to a vertical prefactor). However, the MD $\alpha$ relaxation time is systematically shorter than its experimental DS counterpart by a factor 2 to 3. It is important to note the absolute value of the relaxation time is affected in the simulation by the choice of $\mu$ which was optimized by Blieck~\textit{et al.}~\cite{blieck_molecular_2005} on neutron scattering data for temperatures larger than 333~K (it was already visible that at the smallest temperature simulated by the authors, 313~K, the relaxation time was underestimated). It seems that this is also the case for other studies using the same parameters~\cite{becher2021molecular}. The temperature evolution of the slope $\beta^\mathrm{tot}_{1}$ is also comparable in MD and in experiments (for which it comes from a Cole-Davidson fit to the spectra), as shown by grey solid circles in fig.~\ref{fig6}b but the slopes are systematically underestimated by the simulation.

Contrary to DS, which probes the reorientational dynamics of the molecules by following the permanent dipolar moment, DDLS does it by following the anisotropy of the polarisability tensor. Glycerol molecules are believed to rotate as a rigid entity~\cite{becher2021molecular} and we can reasonably assume that both techniques probe the same dynamics (although giving access to different ranks $\ell$). It is also worth noting that results from DDLS experiments can also be affected by a scattering mechanism called dipole-induced-dipole related to the fluctuation of the internal field. Cummins~\textit{et al.}~\cite{cummins1996origin} have argued that this effect may be neglected in the experimental spectra of supercooled liquids. Here we follow these authors. The spectra $\chi^{\prime\prime~\mathrm{tot}}_{2}(f)$, from the simulation, should therefore be comparable with the Fourier transform of the DDLS correlation function reported by Gabriel~\textit{et al.}~\cite{gabriel_intermolecular_2020}. The authors concentrated mainly on frequencies under a few MHz and thus on temperatures smaller than 260~K (although they also performed a measurement at 323~K). Here also, the MD spectra show a good qualitative agreement with experiments. The DDLS mean relaxation time is shorter than in DS by an amount that appears similar in the experiments and in MD. The inequality $\beta^\mathrm{tot}_{2} < \beta^\mathrm{tot}_{1}$ is also verified for both. In the DDLS experiments (see grey empty circles in fig.~\ref{fig6}d), given the uncertainty, the slope $\beta^\mathrm{tot}_2$ does not appear to depend much on the temperature and it is also the case in the simulation (empty red circles). However, here again, the slopes are underestimated in the simulation.

The differences between the simulation and the real system must be weighed against the relative simplicity of the modeling. Indeed, force field parameters, that control intra and inter-molecular interactions were not adjusted specifically on glycerol but were designed to be applicable to the widest range possible of organic compounds. The only parameter that was adjusted specifically to glycerol was $\mu$. Moreover, the partial charges in the molecule are considered as fixed point charges at the center of each atom, the electronic polarizability is not taken into account and H-bonds are mimicked only by electrostatic interactions of these fixed charges which can limit their strength~\cite{xu2002can}. Nonetheless, it is already impressive that a classical model can reproduce well some aspects of the real system. With these limitations in mind, MD simulations can be used, as demonstrated recently by Becher \textit{et al.}~\cite{becher2021molecular}, to obtain information on microscopic observables or on the relative effect of external parameters such as the temperature.

For $\ell = 2$, and in contrast to the $\ell=1$ case, the effect of cross-correlation is very weak and the total loss function can reasonably be assimilated to the self loss function alone (see fig.~\ref{fig4}d). This is in complete agreement with the observations of Gabriel~\textit{et al.}~\cite{gabriel_intermolecular_2020} and previous work by the same team~\cite{gabriel2017debye, pabst2020dipole} on the ability for DDLS to give access to the self orientational dynamics. This is also in agreement with the recent MD observations on a model system of Koperwas~\textit{et al.}~\cite{koperwas2022computational} who compared the self and total correlation functions for $\ell=2$ at long times ($t \geq \tau_\alpha$).

The ratio of the self relaxation times $\tau^\mathrm{self}_{1}/ \tau^\mathrm{self}_{2}$, shown in fig.~\ref{fig6}c, ranges from 1.4 to 2.0 and is increasing with temperature. This quantity is affected by the details of the molecular relaxation mechanism with two limiting cases leading to values of 3 for isotropic rotational diffusion and 1 for discrete jumps of random amplitude~\cite{bottcher1978dielectrics}. However, a given value of this ratio cannot be associated with a typical orientational jump amplitude. For example, in a mean field model, tuning the interaction parameter allows to change continuously this ratio~\cite{coffey2005nonlinear}. In the present case, it is more likely that its value reflects a dynamics governed by rare relaxation events, that are less averaged for $\ell=2$ than for $\ell=1$~\cite{diezemann1998rotational}. In this case, its decrease upon cooling could be associated with rarer and rarer events. The slope $\beta < 1$ can also be linked with a dynamic governed by relatively rare events, spread on long time scales, that does not average enough to produce an exponential relaxation ($\beta = 1$)~\cite{diezemann1998rotational}. In this context, as $C_{2}$ cancels out for smaller angles than $C_{1}$, it is less averaged over relaxation events leading to $\beta_{2}<\beta_{1}$, consistently with our observations.

In the DDLS works mentioned above~\cite{gabriel2017debye, pabst2020dipole, gabriel_intermolecular_2020}, the authors suggested to use these measurements as a proxy to approach the DS self response, which is not accessible experimentally. This means admitting that the response is reasonably independent of the rank $\ell$ but this also requires a way to normalize the DDLS data so that they can be quantitatively compared to DS spectra. With a well-chosen, temperature independent, scaling factor, it is possible a get a perfect overlap of the DS and DDLS spectra on the excess wing flank at low temperature ($T<200$~K). From this, it was shown that the DS signal could be well fitted by the sum of the DDLS signal and a cross-correlation term (described by a stretched exponential of fixed stretching parameter), with the relative weight of these two terms perfectly matching the Kirkwood correlation factor at all temperatures~\cite{gabriel_intermolecular_2020}. This demonstrates the usefulness and relevance of the direct comparison between DDLS and DS spectra.

In our numerical work, we were not able to reach the range of low temperatures in which the excess wing appears. This prevented us from using the method described above to determine from scratch a scaling factor between the $\ell = 1$ and $\ell = 2$ spectra. We instead had to rely on the experimental determination by Gabriel~\textit{et al.}~\cite{gabriel_intermolecular_2020}. Indeed, while the DS and DDLS spectra collapse on the excess wing for $T<200$~K, they cross at a finite frequency $f_\mathrm{cross}$ for higher temperatures. We determined a ($T$ independent) scaling factor by making sure that the $\ell = 1$ and $\ell = 2$ total spectra cross at the same $f_\mathrm{cross}/f_\alpha$ than the experiments for $T = 323$~K and $T = 263$~K (see vertical dashed lines in fig.~\ref{fig5}c). The result is shown in fig.~\ref{fig5}c and it appears that, for all temperatures, the $\alpha$ peak of the total $\ell = 2$ and the self $\ell = 1$ spectra match fairly well. The difference of slopes $\beta$ is small enough so that the discrepancy remains low ($< 30$~\%) under the frequency at which the fast process starts to be perceptible. This allows us to understand the success of the experimental approach consisting of a direct comparison between DS and DDLS data~\cite{gabriel2017debye, pabst2020dipole, gabriel_intermolecular_2020}.

\section{Conclusion and perspectives.}

In this work, we studied using MD simulations, the orientational dynamics of glycerol from which we extracted the self correlation function and the associated loss function for different ranks $\ell$ of the Legendre polynomial. For $\ell = 1$ and 2, we studied the spatial dependence of the cross-correlation and showed that they play a significant role in the $\ell = 1$ response while being almost negligible for $\ell = 2$. In accordance with recent experimental observations based on a comparison between DDLS and DS spectra, we showed that, although these techniques give access to different ranks $\ell$ of the correlation function (and consequently do not lead to the exact same spectra, in particular regarding the slope $\beta$), the scaling factor that corresponds to a merging of the excess wing of DS and DDLS spectra at low temperatures leads to a fairly good merging of the self part of the $\ell=1$ and the total $\ell = 2$ spectra. This strengthens the idea that useful information can arise from a direct comparison between DDLS and DS measurements. 

Moreover, we took advantage of the possibility given by MD simulations to access molecular observables to discuss the microscopic origin of the cross-correlation observed in DS. We found out that the net cross-correlation originates from the first shell of neighbouring molecules that tend to positively align independently of their orientation. Investigating in more detail the molecular origin of the cross-correlations, and their link with molecular conformation and H-bonds will be the subject of future work.

\begin{acknowledgments}
We thank J.P. Gabriel for sharing the data published in ref.~\cite{gabriel_intermolecular_2020} and C. Alba-Simionesco for discussions. M.H is grateful to LABEX PALM and IRAMIS Institute for ﬁnancial support. This work was supported by ANR PIA funding: ANR-20-IDEES-0002.
\end{acknowledgments}


%

\clearpage
\onecolumngrid
\renewcommand\thefigure{S\arabic{figure}}    
\renewcommand\thetable{S\arabic{table}}    

\renewcommand{\theequation}{S\arabic{equation}}
\setcounter{equation}{0}
\setcounter{figure}{0}
\setcounter{table}{0}
\setcounter{section}{0}

\begin{center}
   
\large \textbf{Supplementary materials}
\end{center}

\section{Force field for glycerol}

The force field used in this work is the one initially introduced by Chelli~\textit{et al.}~\cite{chelli1999glycerol, chelli_glycerol_1999_2}, later optimized by Blieck~\textit{et al.}~\cite{blieck_molecular_2005} and used since with only little differences by Egorov~\textit{et al.}~\cite{egorov_molecular_2011}, Busselez~\textit{et al.}~\cite{busselez2014structural} and Becher~\textit{et al.}~\cite{becher2021molecular}. The partial charges used for the glycerol molecules are shown in fig.~\ref{figures_suppl_charges}. The harmonic bound, harmonic angle, and periodic torsion forces parameters are given in table~\ref{tableSuppl_bond}, \ref{tableSuppl_angle} and \ref{tableSuppl_torsion} respectively. The Lennard-Jones parameters are given in table~\ref{tableSuppl_LJ}.

\section{Simulation runs}

The simulation runs used in this work are described in table~\ref{tableSuppl_sim}. At each temperature $T<323~K$, the equilibration is performed starting from the equilibrated state at $T + 10$~K.

\section{Effect of the simulation box size}

\begin{figure*}[htbp]
  \centering
  \includegraphics[width=\linewidth]{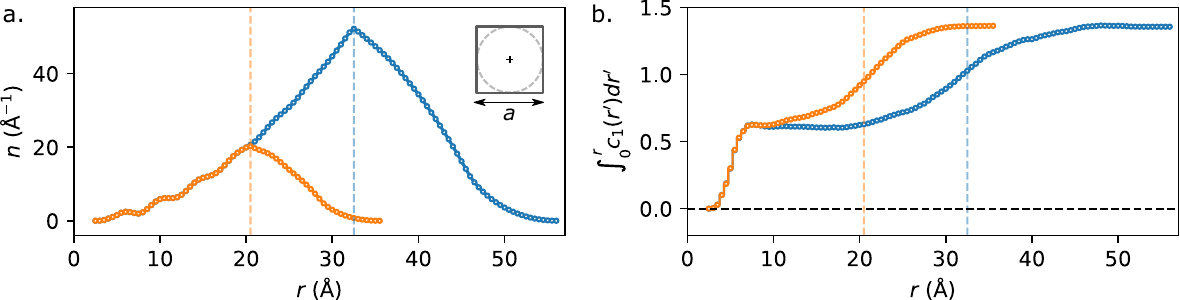}
 \caption{Effect of the simulation box size at $T = 323$~K $N = 2160$ molecules used in this study (blue) and $N = 540$ (orange). (a) Dipole density $n(t)$. The vertical dashed lines correspond to the half side $a/2$ of the simulation box. (b) Integrated cross-correlation $\int_0^r c_1(r', t=0) \mathrm{d}r$.}
  \label{suppl_fig_box}
\end{figure*}

The effect of the simulation box size was studied by performing an additional simulation run at 323~K for only $N=540$ molecules (7560 atoms) in a cubic cell of side length $a\approx 41$ ~\AA. The dipole density $n(r)$ and the integrated cross-correlation contribution $\int_0^r c_1(r', t=0) \mathrm{d}r$ are shown in fig.~\ref{suppl_fig_box} for this case as well as for the box size used in the main text ($N=2160$, $a \approx 65$~\AA) at the same temperature. Both datasets display a similar behavior regarding cross-correlations: a first plateau is reached at $r~\approx 8$ ~\AA followed by an increase with a maximum slope reached at $r = a/2$. This box-size dependent second regime is due to the effect of PBCs on the treatment of electrostatic interactions with the PME method~\cite{caillol1992asymptotic, zhang2016computing, olivieri2021confined}. Using $N = 2160$ allows us to safely decouple the PCBs artifact from the physically meaningful behavior at $r< 15$ ~\AA.

\section{Comparison of $\mu^2g_\mathrm{K}$ with experiments}
\begin{figure}[htbp]
  \centering
  \includegraphics[width=6cm]{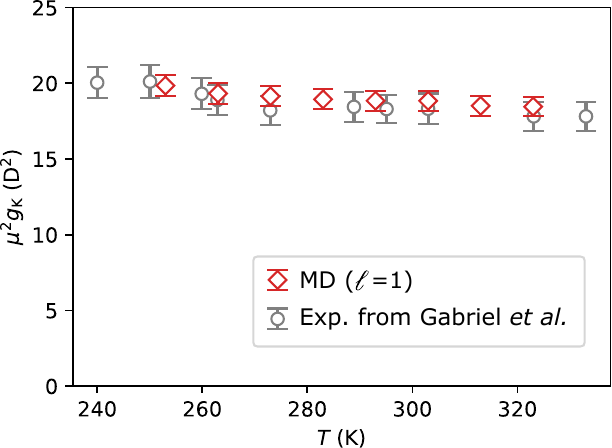}
 \caption{Temperature dependence of $\mu^2 g_\mathrm{K}$ obtained from the MD simulation and experimentally from the measurement of the dielectric strength (from ref.~\cite{gabriel_intermolecular_2020}).}
  \label{suppl_mu2_gk}
\end{figure}

\begin{figure}[htbp]
  \centering
  \includegraphics[width=7.5cm]{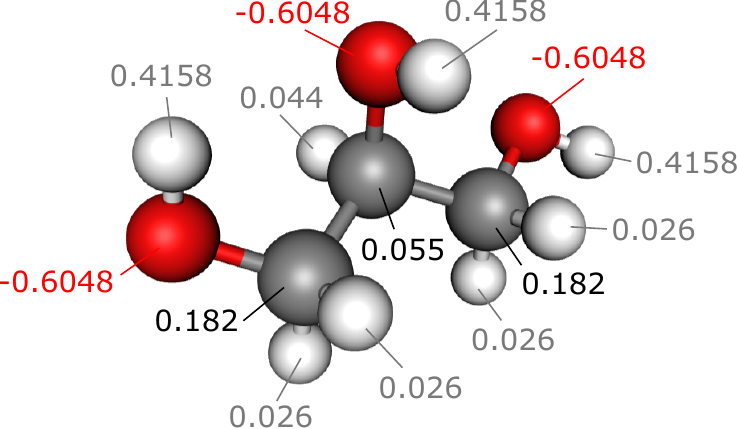}
 \caption{Atomic charges of the glycerol molecule.}
  \label{figures_suppl_charges}
\end{figure}

Fig.~\ref{suppl_mu2_gk} shows $\mu^2 g_\mathrm{K}$ obtained from the MD simulation in this work (red) and from experimental measurements of $\epsilon(0)$ and eq.~\ref{KKVRSeq} by Gabriel~\textit{et al.}~\cite{gabriel_intermolecular_2020}.

\begin{table}
\caption{Harmonic bond force parameters. The potential is of the form $E_\mathrm{b} = k_\mathrm{b} (r-r_0)^2$.}
\label{tableSuppl_bond}
\begin{tabularx}{0.5\textwidth}{c 
>{\centering\arraybackslash}X
>{\centering\arraybackslash}X}
\hline
  bond & $k_\mathrm{b}$ (kcal$\cdot$mol$^{-1}\cdot$\AA$^{-2}$) & $r_0$ (\AA) \\
  \hline
  \hline
   CC & 310 & 1.526  \\
   CO & 320 &  1.410 \\
   CH & rigid bond &  1.090 \\
   OH & rigid bond &  0.960 \\
\hline
\end{tabularx}
\end{table}

\begin{table}
\caption{Harmonic angle force parameters. The potential is of the form $E_\theta = k_\mathrm{\theta} (\theta-\theta_0)^2$.}
\label{tableSuppl_angle}
\begin{tabularx}{0.5\textwidth}{c 
>{\centering\arraybackslash}X
>{\centering\arraybackslash}X}
\hline
  angle & $k_\mathrm{\theta}$ (kcal$\cdot$mol$^{-1}\cdot$rad$^{-2}$) & $\theta_0$ (°) \\
  \hline
  \hline
   CCC & 40 & 109.5  \\
   CCO & 50 &  109.5 \\
   CCH & 50 &  109.5 \\
   COH & 55 &  108.5 \\
   HCH & 35 &  109.5 \\
   OCH & 50 &  109.5 \\
\hline
\end{tabularx}
\end{table}

\begin{table}
\caption{Periodic torsion force parameters. The potential is of the form $E_\Phi = k_\mathrm{\Phi} (1+\cos(n\Phi))$.}
\label{tableSuppl_torsion}
\begin{tabularx}{0.5\textwidth}{c 
>{\centering\arraybackslash}X
>{\centering\arraybackslash}X}
\hline
  dihedral & $V_\mathrm{n}$ (kcal$\cdot$mol$^{-1}$) & $n$ \\
  \hline
  \hline
CCCH & 0.1556 & 3 \\
CCCO & 0.1556 & 3 \\
OCCO & 0.1440 & 3 \\
OCCO & 1.0000 & 2 \\
OCCH & 0.1556 & 3 \\
HOCC & 0.1667 & 3 \\
HOCH & 0.1667 & 3 \\
HCCH & 0.1556 & 3 \\
\hline
\end{tabularx}
\end{table}

\begin{table}
\caption{Lennard-Jones parameters. The potential is of the form $V_\mathrm{LJ} = 4 \left( \left(\frac{\sigma}{r}\right)^{12}-\left(\frac{\sigma}{r}\right)^{6}\right)$. The input parameters of the Amber .frcmod file is the half atom-atom distance at which the potential reaches its minimum ($R_\mathrm{m} = 2^{1/6}\sigma$).}
\label{tableSuppl_LJ}
\begin{tabularx}{0.5\textwidth}{c 
>{\centering\arraybackslash}X
>{\centering\arraybackslash}X
>{\centering\arraybackslash}X}
\hline
  atom & $R_\mathrm{m}/2$ (\AA) & $\sigma$ (\AA) & $\epsilon$ (kcal$\cdot$mol$^{-1}$) \\
  \hline
  \hline
C & 2.1416 &  3.816 & 0.1094 \\
O & 1.6000 & 2.850 & 0.1591 \\
aliphatic H & 1.5569 & 2.774 & 0.0157 \\
hydroxyl H & 0.800 & 1.425 & 0.0498 \\
\hline
\end{tabularx}
\end{table}

\begin{table*}[t]
\centering
\caption{Details on the simulation runs.}
\label{tableSuppl_sim}
\begin{tabularx}{\textwidth}{c 
>{\centering\arraybackslash}X
>{\centering\arraybackslash}X
>{\centering\arraybackslash}X
>{\centering\arraybackslash}X
>{\centering\arraybackslash}X
>{\centering\arraybackslash}X
}
\hline
  $T$ (K) & equil. steps & in $\tau_\mathrm{self}$ unit. & in s & run steps & in $\tau_\mathrm{self}$ unit.& in s \\
  \hline
  \hline
323 & $4\times 10^6$ & 213 & 16 ns &  $3.5\times 10^6$ & 186 & 14 ns\\
313 & $7\times 10^6$ & 225 & 28 ns & $7\times 10^6$ & 225 & 28 ns\\
303 & $14\times 10^6$ & 253 & 56 ns & $14\times 10^6$ & 253 & 56 ns\\
293 & $33\times 10^6$ & 262 & 130 ns & $30\times 10^6$ & 238 & 120 ns \\
283 & $38\times 10^6$ & 114 & 150 ns & $75\times 10^6$ & 225 & 300 ns \\
273 & $108\times 10^6$ & 113 & 430 ns & $215\times 10^6$ & 226 & 430 ns\\
263 & $358\times 10^6$ & 98 & 1.4 $\upmu$s & $710\times 10^6$ & 194 & 2.8  $\upmu$s\\
253 & $1830\times 10^6$ & 101 & 7.3 $\upmu$s & $1210\times 10^6$ & 67 & 4.8  $\upmu$s\\
\hline
\end{tabularx}
\end{table*}

\end{document}